\documentstyle{mn}
\title{Heavy neutrino ball as a possible solution to the
 "blackness problem" of the Galactic center}
\author[David Tsiklauri and Raoul D. Viollier]
       { David Tsiklauri and Raoul D. Viollier\\
        Physics Department, University of Cape Town,
Rondebosch 7701, South Africa}
\date{Accepted ????.
      Received ????;
      in original form ????}
\pagerange{\pageref{firstpage}--\pageref{lastpage}}
\pubyear{1998}
\begin{document}
\maketitle

\begin{abstract}
It has been recently shown (Tsiklauri \& Viollier, 1998a)
that the matter concentration inferred from observed
stellar motion at the galactic center
(Eckart \& Genzel, 1997, MNRAS, 284, 576 and
Genzel et al., 1996, ApJ, 472, 153) is consistent with a supermassive
object of $2.5 \times 10^6$  solar masses, composed of
self-gravitating, degenerate heavy neutrinos.
It has been furthermore suggested
 (Tsiklauri \& Viollier, 1998a) that the neutrino ball
scenario may have an advantage
that it could possibly explain the so-called
"blackness problem" of the galactic center. Here, we present
a quantitative investigation of this statement, by calculating  the
emitted spectrum of Sgr A$^*$ in the framework of standard accretion
disk theory.
\end{abstract}


\section{Introduction}
The enigmatic radio source Sgr A$^*$ at the galactic center
has been a longstanding puzzle. Observations of
stellar motions at the galactic center
(Eckart \& Genzel, 1997; Genzel et al., 1996)
and low proper motion ($\leq$ 20 km sec$^{-1}$; Backer, 1996) 
of  Sgr A$^*$
indicate that, on the one hand,
it is a massive $(2.5 \pm 0.4) \times 10^6 M_\odot$
object dominating the gravitational potential in the 
inner $\leq 0.5$ pc
region of the galaxy. On the other hand,
observations of stellar winds and other gas flows in the vicinity
of Sgr A$^*$
suggest that the mass accretion rate $\dot M$ is about $6 \times
10^{-6}M_\odot$yr$^{-1}$ (Genzel et al., 1994). This implies that
the luminosity
of the central object should be more than $10^{40}$ erg sec$^{-1}$, provided
the radiative efficiency is the customary 10\%. 
However, observations indicate that the
bolometric luminosity is actually less than $10^{37}$ erg sec$^{-1}$.
This discrepancy has been a source of exhaustive debate in the 
recent past.
The broad-band emission spectrum of  Sgr A$^*$
can be reproduced either in the quasi-spherical accretion model
(Melia, 1992, 1994) with $\dot M \simeq 2 \times 10^{-4}M_\odot$ 
yr$^{-1}$ or by
a combination of disk plus radio-jet model 
(Falcke et al., 1993a, 1993b).
As pointed out by Falcke and Melia (1997), quasi-spherical accretion
seems  unavoidable at large radii, but the low actual luminosity
of Sgr A$^*$ points toward a much lower 
accretion rate in a starving disk.
Therefore, Sgr A$^*$ can be described by a model of a
 fossil disk fed by quasi-spherical accretion.
Another successful model which is consistent with the 
observed emission
spectrum of Sgr A$^*$ has been developed
by Narayan et al., 1995, 1998 and
independently by Manmoto et al., 1997. This model is
based on the concept of advection dominated accretion flow, in which
 most of the energy released by viscosity in the disk 
is carried along
with the gas and lost into the black hole, while 
only a small fraction is actually radiated off.

Recently, Tsiklauri \& Viollier  (1998a) have proposed
an alternative model for the mass distribution at the galactic
center in which the customary supermassive 
black hole is replaced by
 a ball composed of self-gravitating, degenerate neutrinos.
It was shown that
a neutrino ball with a mass $2.5 \times 10^6 M_\odot$,
composed of neutrinos and antineutrinos with  masses
$m_\nu \geq 12.0$ keV$/c^2$
for $g=2$ or $m_\nu \geq 14.3$ keV$/c^2$ for $g=1$, where $g$
is the spin degeneracy factor, is  consistent with the
current observational data.  The purpose of this paper is to
present
calculations of the spectrum emitted by Sgr A$^*$ in the framework of
standard accretion disk theory, assuming that 
Sgr A$^*$ is a neutrino
ball with the abovementioned physical properties, and to show
that this could resolve the "blackness problem".

In the recent past, Viollier et al. have proposed that
massive, self-gravitating, degenerate 
neutrinos arranged in balls, where
the degeneracy pressure balances self-gravity, can
form long-lived configurations that could mimic
the properties of dark matter
at the centers of galaxies
(Viollier, 1994; Viollier et al., 1993; Viollier et al., 1992).
Tsiklauri \& Viollier (1996)  have shown that a neutrino ball
could play a similar role as a stellar cluster in the 3C 273 quasar,
revealing its presence through the infrared bump in the emitted
spectrum.
Tsiklauri \& Viollier (1998b) further investigated the
formation  and time evolution of neutrino balls via
two competing processes:
annihilation of the particle-antiparticle pairs via weak interaction
and spherical (Bondi) accretion of these particles.
Bili\'c \& Viollier (1997) showed how the neutrino balls could
form via a first-order phase transition
of a system of self-gravitating neutrinos in the presence
of a large radiation density background, based on the
Thomas-Fermi model at finite temperature. They find
that, by cooling a non-degenerate gas of massive
neutrinos below a certain critical temperature, a condensed phase
emerges, consisting of quasi-degenerate supermassive neutrino
balls.
General relativistic effects in the study of the gravitational phase
transition in the framework of the 
Thomas-Fermi model at finite temperature
were taken into account in Bili\'c \& Viollier (1998a).
A theorem was proven by Bili\'c \&
Viollier (1998b) which in brief states that the extremization of the
free energy functional of a system of self-gravitating fermions,
described by the general relativistic 
Thomas-Fermi model, is equivalent
to solving Einstein's field equations.

\section{The model}

The basic equations which govern the structure of
cold  neutrino balls have been derived in the series of papers
(Viollier, 1994; Viollier et al., 1993; Viollier et al., 1992
and Tsiklauri \& Viollier, 1996); here
we adopt the notation of Tsiklauri and  Viollier (1996). In
this notation the
enclosed mass of the neutrinos and antineutrinos within a radius
$r=r_n \xi$
of a neutrino ball is given by
$$
M_<= 8 \pi \rho_c r_n^3  
\left({-\xi^2 {{d \theta(\xi)}\over{d \xi}} }\right)
\equiv 8 \pi \rho_c r_n^3 \left({-\xi^2 \theta^{\prime}}\right), \eqno(1)
$$
where $\theta (\xi)$ is the standard solution of the Lane-Emden
equation with polytropic index $3/2$, $r_n$ is the Lane-Emden unit of
length
and $\rho_c$ is the central density of the neutrino ball.

In the standard theory of steady and geometrically thin accretion disks,
the power liberated in the disk per unit area is given by
(Perry \& Williams, 1993)
$$
D(r)=-{{\dot M \Omega \Omega^\prime r}\over{4 \pi}} \left[
1-\left({R_i \over r}\right)^2 \left({\Omega_i \over \Omega}
\right) \right]. \eqno(2)
$$
Here $\Omega$ is the angular velocity of the accreting matter,
$R_i$ is the inner edge of the disk and $\Omega_i$ denotes  the
angular velocity at the radius where its derivative with respect to $r$
vanishes due to the deviation from the Keplerian law of rotation.
Finally, the prime denotes the derivative with respect to $r$. Since
the motion of  accreting matter in the bulk of the disk is
Keplerian, we assume that the angular velocity is given by
$$
\Omega(r)=\sqrt{{G M_<(r) \over r^3}}. \eqno(3)
$$
In the case of a back 
hole $M_<(r)=const=M_{bh}$, whereas in our case $M_<(r)$
is determined by Eq.(1). 
Throughout this paper we take
 the outer radius of the disk as
 $10^5$ Schwarzschild radii, since for
 larger radii the disk is  unstable against  
self-gravity (e.g. Narayan et al., 1998). The
radius of a neutrino ball with a mass $2.5 \times 10^6 M_\odot$,
composed of neutrinos and antineutrinos with  masses
$m_\nu = 12.0$ keV$/c^2$
for $g=2$ or $m_\nu = 14.3$ keV$/c^2$ for $g=1$, is equal to
1.06 $\times 10^5$ Schwarzschild radii of a black hole with 
the same mass,
thus the accretion disk is fully immersed in the neutrino ball.
Moreover, 
as in our case there is no last stable orbit, 
 accretion may in principle continue as $r$ tends to zero, where
$\Omega(r)$ and $\Omega^\prime(r)$ assume the values
$$
\Omega(0)=\sqrt{{8 \pi G \rho_c} \over 3}, \;\;\;\;\;\;\;\;
\Omega^\prime(0)={{1.5 \pi G \rho_c}\over{r_n \Omega(0)}}. \eqno(4)
$$
Of course the latter result is of rather academic interest, because
in reality, the accreting matter will be diverted at the origin
in the form of
an outflow which will inevitably stream away perpendicular to the
disk plane.  The excess matter which has spiraled
down to the very center will be pushed out of the plane
due to the gas pressure of the accreting matter in the 
disk. It is important to note that this outflow will differ
considerably from a jet shooting out of an accretion disk
around a black hole. In the latter case, the jets manifest
themselves as strong emitters mostly in the radio band
due to the synchrotron radiation produced by the electrons
moving at highly relativistic velocities,
whereas the  outflow from the accretion disk
immersed in a neutrino ball will be practically
unobservable, since the outflowing matter will be cold as
it radiated off its energy while
spiraling down in the disk (see further Fig.2).
Moreover, the particles 
 will be moving at non-relativistic
velocities because of the  shallowness of the 
gravitational potential of the  neutrino ball that is much more
spatially extended than a black hole.
Also, it is worthwhile to note that, even at a constant
accretion rate of $6 \times 10^{-6} M_\odot$ yr$^{-1}$,
the baryonic mass acquired by the neutrino ball
within the age of the universe of 10 Gyr would be 
of the order of
$6 \times 10^4 M_\odot$ which is small compared to
the mass of the neutrino ball.

Numerical analysis shows that initially, as the matter spirals
towards the center,  $\Omega^\prime(r)$ is negative. From  Eq.(4)
we gather that the central value for $\Omega^\prime$ is finite and
positive, thus there exits a point at which $\Omega^\prime$ crosses
zero. This is precisely the point where the 
angular velocity attains its
maximal value. Numerically, this happens 
at $\xi_i= R_i/r_n=8.25 \times 10^{-4}$.
Note that this position is quite close to the center of the ball
since its radius in dimensionless units is $\xi_1=3.65375$
(Cox \& Giuli, 1968).
Such a behavior of $\Omega(r)$ is quite interesting since, in
the neutrino ball scenario,
 there is neither  last stable orbit (as in the case of a black hole)
nor  a stiff stellar surface 
(as in the case of accretion onto a neutron star).
Basically, this is a consequence of the non-trivial mass distribution
determined by the Lane-Emden equation.

We now assume that the gravitational binding energy
released is immediately radiated away locally according to the
 Stefan-Boltzmann's law
$$
D(r)=\sigma T^4_{\rm eff}(r), \eqno(5)
$$
with $\sigma$ denoting Stefan-Boltzmann constant.
The effective temperature can be derived  using Eqs.(1-3) and (5)
yielding
$$
T_{\rm eff}(\xi)=\left[- {{\dot M \bar \Omega \bar \Omega^\prime
r_n}\over{4 \pi \sigma}}
{{(\xi \theta^{1.5}+3 \theta^\prime)}\over{\xi}}
\left(1- \xi_i^2 \bar \Omega_i 
\sqrt{-{1\over{\theta^\prime \xi^3}}}\right)
\right]^{1/4}. \eqno(6)
$$
Here we have introduced the quantities
$$
\bar \Omega =\sqrt{{2.5 \times 10^6M_\odot G}\over{r_n^3
(-\xi^2\theta^\prime)_1}}, \;\;\;\;\;\;
 \bar \Omega^\prime={{2.5 \times 10^6 M_\odot G}\over{2 \bar \Omega
 r_n^4 (-\xi^2\theta^\prime)_1}},
$$
 $(-\xi^2\theta^\prime)_1=2.71406$ (Cox \& Giuli, 1968)
and $\bar \Omega_i = \Omega(\xi_i)/ \bar \Omega$.

Once the temperature distribution in the accretion disk is specified,
we may calculate  its luminosity using 
$$
L_\nu={{16 \pi^2 h r_n^2\cos i \nu^3}\over{c^2}}
\int_{\xi_i}^{\xi_1} 
{{\xi d \xi}\over{\exp[h \nu/k_{\rm b} T(\xi)]-1}},
 \eqno(7)
$$
where $h$ is Planck's constant, $k_{\rm b}$ denotes Boltzmann's constant
and $i$ is the disk inclination angle which we assume to be 
 $60^\circ$ as in 
 Narayan et al. (1998). Following the same paper,
we parameterize the accretion rate
in terms of the Eddington limit accretion rate, i.e.
$\dot M= \dot m \dot M_{\rm Edd} M_\odot$yr$^{-1}$,
where $\dot M_{\rm Edd}=10 L_{\rm Edd}/c^2=1.39 \times 10^{18}
(M/M_\odot)$g sec$^{-1}$=$2.21 \times 
10^{-8}(M/M_\odot) M_\odot$yr$^{-1}$.
Melia (1992) has estimated $\dot M$ as 
$\approx 2 \times 10^{-4}M_\odot$yr$^{-1}$
using 600 km sec$^{-1}$ for the wind velocity, 
whereas Genzel et al. (1994)
obtained $\dot M \approx 6 \times 10^{-6}M_\odot$yr$^{-1}$ using
1000 km sec$^{-1}$ for the wind velocity. These values translate into
$10^{-4}< \dot m < 4 \times 10^{-3}$ in terms of the Eddington units.
Following again Narayan et al., 1998 we use 
these two  values as the lower and
upper limits for this quantity. 

\section{Discussion}

Results of our numerical calculations are presented
in  Fig.1, where we plot the quantity $\nu L_\nu$, calculated
using Eq.(7). Data points are taken from
Table 1 in Narayan et al., 1998.
The thick solid line corresponds to the case of a neutrino ball with
$\dot m=4 \times 10^{-3}$, whereas the
thin solid line corresponds to $\dot m= 10^{-4}$. 
The short-dashed line
represents  the calculation with a 
$2.5 \times 10^6 M_\odot$  black hole
with $\dot m= 10^{-4}$ and an accretion disk 
extending from from 3 to
$10^5$ Schwarzschild radii. The long-dashed 
line corresponds to the case
when $\dot m$ is artificially brought down to $10^{-9}$. 
As we see from  Fig.1,
and as also was pointed out by Narayan et al., 1998, the latter two
curves provide a  poor fit to the observational data.  Actually,
this is
the major reason why the standard  accretion disk theory was abandoned
as a possible candidate for the description of 
the emitted spectrum from Sgr A$^*$.
However, as originally was pointed out in Tsiklauri \& Viollier (1998a)
 in the neutrino ball scenario,
the accreting matter  experiences
a much shallower gravitational potential than in the case of
the black hole with the same mass, and
therefore less viscous torque will be exerted.
The radius of a neutrino ball of total mass
$2.5 \times 10^6 M_\odot$,
which is composed of self-gravitating,
degenerate neutrinos and antineutrinos of mass  $m_\nu = 12.0$ keV$/c^2$
for $g=2$ or $m_\nu = 14.3$ keV$/c^2$ for $g=1$,  
is $1.06 \times 10^5$ larger
than the Schwarzschild radius of a black hole of the same
mass. In this context it is important to note that the accretion
radius $R_{\rm A}=2GM/v^2_{\rm w}$ for the neutrino ball,
where $v_{\rm w}\simeq 700$ km/sec is the velocity of the wind from
the IRS 16
stars, is approximately 0.02pc (Coker \& Melia, 1997), which is slightly
less than the radius of the neutrino ball, i.e. 0.02545 pc
(for $m_\nu = 12.0$ keV$/c^2$
for $g=2$ or $m_\nu = 14.3$ keV$/c^2$ for $g=1$).  
Therefore, in
the neutrino ball scenario, the captured accreting matter will
always experience a
gravitational pull from a mass less than the total mass of the ball.
One can see from  Fig. 1, that for this very reason the theoretical
spectrum in the case of the neutrino ball with 
$\dot m=4 \times 10^{-3}$ gives a much better fit than in the case
of a black hole for any (even unrealistically lowered) values of
$\dot m$.
Discrepancies between the theoretical and observed spectra appear
in the case of the neutrino ball  for frequencies
$<40$ GHz and $\geq 10^{14}$ Hz.

At the higher end ($\geq 10^{14}$ Hz) of the spectrum,
the discrepancy is due to the fact that  our model does not 
incorporate effects of Compton-scattered synchrotron radiation
(which  causes the second peak on the left in  Fig.1 of
Narayan et al., 1998).  Our model is based on the simple-minded
assumption of a 
steady, geometrically thin accretion disk which radiates
off the gravitational binding energy locally, 
according to the black-body radiation law.
 However, even in this simplified framework, our
model gives a reasonable fit in the radio to 
near infrared part of the
spectrum. Besides, it is important to note that,
as it has been shown by
Falcke \& Melia (1997), the evolution of an accretion
disk can be considerably influenced by the deposition of mass and angular
momentum by an infalling Bondi-Hoyle wind. The major result of their
paper is that the modification of the standard
accretion disk model, by taking into account the contribution from the
Bondi-Hoyle wind and considering the physical picture of accretion
process in dynamics, yields significant changes in the emitted spectrum.
In fact, it produces an infrared bump, in addition to the Big Blue Bump,
due to the deposition of energy in the outer part of the fossil accretion
disk. Our paper is based on the standard accretion disk model i.e.
without modifications arising from taking into account effects
from the wind. In our case the gravitational potential is shallower
than in the case of a supermassive black hole with the same mass.
Therefore, taking into account effects from the Bondi-Hoyle wind and
considering the non-steady problem (as in the case of Falcke \& Melia's
paper), both bumps will be shifted into the lower frequency domain.
Thus the incorporation of Falcke \& Melia's model of the accreting flow
into our scenario of the dark matter distribution at the galactic center
would presumably produce a better fit in the $\leq 40$ GHz part of the
spectrum.

It is important to address the issue of consistency of our
model with intrinsic source size versus frequency
data. For the test we take the  data
 of emission wavelength $\lambda =$ 7 mm (Bower \& Backer, 1998)
and 3.5 mm (Rogers et al., 1994; Krichbaum et al., 1994). 
The upper limits on the
intrinsic source size are $< 4.1$ AU (Bower \& Backer, 1998)
 for 7 mm 
and $< 1.1$ AU (Rogers et al., 1994) and 
$2.8 \pm 1.2$ AU (Krichbaum et al., 1994) for 3.5 mm assuming
a distance to the galactic center of 8.5 kpc.
Now, we have to estimate the radial location of the
circles of the accretion disk in our model, emitting at 
these two wavelengths. For this purpose we assume that 
the corresponding temperature of a circle can be determined by the 
Wien displacement law $\nu_m \approx 3 k_{\rm b} T /h \approx 6 \times 
10^{10} T$ Hz (Lang, 1974), i.e.  we assume that
the maximal frequency (wavelength) in the brightness
distribution given the black body law determines
the temperature of the emitting region. This assumption
seems reasonable recalling the sharpness of the maxima
in the brightness distribution (see Fig. 1 in Lang, 1974).
Therefore we obtain $T_{\rm 7mm}=0.71 K$ and $T_{\rm 3.5mm}=1.43 K$.
To find out to which values of the radial distance in the accretion disk
these two values correspond, we have to use Eq.(6), graphically
depicted in the Fig.2. The values are $\xi_{\rm 7mm}=8.71 \times 10^{-4}$
and $\xi_{\rm 3.5mm}=1.145 \times 10^{-3}$. 
The final predictions of our model would be twice these values
({\it diameter} of the emitting circle) which in dimensional units
are 2.50 AU and 3.29 AU for 7 mm and 3.5 mm, respectively.
Thus we conclude that for 7mm  our model is
consistent with the observations by Bower \& Backer (1998);
the same applies to the  data by Krichbaum et al. (1994) 
for 3.5 mm.
However, currently some of the 
VLBI observations at millimeter wavelength stand in conflict 
with each other (Bower \& Backer, 1998). Therefore, the discrepancy 
of our estimates with the
Rogers et al., (1994) data is a matter of debate.
Another important requirement which our model does satisfy is
the lower limit on the size derived from the scintillation experiments
(Gwinn et al., 1991). These experiments imply that the source
diameter should be $> 0.1$ AU for 0.8 mm wavelength.
Our estimates  show that at this wavelength the
source diameter is 34.11 AU which corroborates the
validity of our model. 

As we can conclude from the latter paragraph, our model
satisfies the source versus frequency constraints. 
However, it seems
unlikely that such low temperatures as required by our model,
especially close to the center of the neutrino ball, are actually
realized, as the Galactic center is immersed in a hot radiation field. 
So far, the lowest termperature of one of the (black body radiation) 
emitting components
introduced by Zylka, Mezger \& Lesh 1992, was estimated 
to be 30 K, as a possible alternative to the self-absorbed
synchrotron emission.
Thus, we do not claim that in the
mm wave-length band (which corresponds to disk temperatures of
order few Kelvin) our model is capable of explaning the emitted
spectrum reliably. 
The fit of the spectrum at these frequences ($<40$ GHz),
based on standard accretion disk theory for a disk immersed in the
field of the neutrino ball, is anyway not good, and
it seems that some other, presumably non-thermal mechanism 
is responsible for the radio  emission of Sgr A$^*$.
Moreover, as  has been shown by Reynolds \& McKee 1980,
the compact radio source  at the Galactic center 
could be a pulsar,
with total emitting power comparable to that of the Crab pulsar.
These authors showed that Sgr A$^*$ can be understood
in the framework of
a number of dynamically self-consistent models of
incoherent synchrotron sources which are energetically 
comparable with to the energy output of a 
few $\times 10^{38}$ erg sec$^{-1}$ like the Crab pulsar.
The radio pulsar, together with  a Shakura-Sunyaev disk embedded
in the shallow gravitational potential of the neutrino ball,
may be the key to the understanding  
of the emission spectrum of Sgr A$^*$.
In this context, it is perhaps important to point out
that the
low proper motion of the central radio source 
($\leq$ 20 km sec$^{-1}$; Backer, 1996) could be explained
by a slowly moving pulsar near the minimum of the
gravitational potential of the neutrino ball, a possibility
which of course would be excluded in a supermassive
black hole scenario.

Apart from emission of high-energy radiation by the
pulsar, and X-ray emission of the neutrino ball
through radiative decay of the constituent neutrinos
into light neutrinos, our model would be incapable
of describing of X-rays and gamma-rays. First, even
standard accretion disk theory around a central black hole
cannot account for the emission of radiation above ultra-violet. 
The simplest model
which might produce X-rays is the two-component plasma
model by Shapiro, Lightman \& Eardley 1976.
However, in the case of a neutrino ball with the physical
parameters mentioned above, it is impossible to get
X-rays and gamma-rays from the accretion disk by {\it definition}. 
In order
to get an appreciable fraction of the rest mass of the 
electron converted into X-rays,
the particles need to reach a sizable fraction of the velocity
of light, which would be impossible in our
scenario as  the escape velocity from the center of the
neutrino ball is about 1400 km s$^{-1}$.
This is, in fact, the reason why our model can explain the
blackness
problem of the Galactic center.

Let us now take an 
unprejudiced view on the high-energy data:
In the 0.8-2.5 keV band the data  avaliable from ROSAT
(Predehl \& Tr\"umper 1994) have a resolution of $\sim 20^{''}$;
in the 2-10 keV band data available 
from  ASCA (Koyama et al. 1996) the resolution is
$\sim 1'$.
The 35-150 keV data from  SIGMA (Goldwrum et al. 1994) have a
resolution of
$\sim 15'$, while the 
EGRET data (Merck et al. 1996) from 30 MeV to 10 GeV 
have a resolution of $\sim 1^\circ$.
The intrinsic size of the accretion disk (which is about 
the size of the neutrino ball) is about $0.65''$ which
is much smaller than the resolution of current
X-ray or gamma-ray detectors. 
Even the measurement by
Predehl \& Tr\"umper 1994, who established that the X-ray source
is within a $10''$ (0.5 pc) distance from the Sgr A$^*$
is not conclusive.
As discussed in detail in Tsiklauri \& Viollier
1998a, a neutrino ball would also produce  X-rays  
via the radiative decay of the 
heavy into light neutrinos. Tsiklauri \& Viollier 1998a
have estimated that the luminosity of this emission
line at energy $\sim m_\nu c^2/2$ (which would not be
a sharp line due to scattering by the existing matter at the
Galactic center) should be $L_\gamma \leq 1.45 \times 10^{34}$ erg/sec.
In fact, this luminosity is consistent with the observations by
Predehl \& Tr\"umper 1994.

Finally, we would like to emphasize that the idea that Sgr A$^*$
may be an extended object rather than a supermassive black hole
is not new (see e.g. Haller et al., 1996; Sanders, 1992).
To our knowledge all 
previous such  models  assume that the extended
object is of a baryonic nature, e.g.  a very compact stellar
cluster. However, it is commonly accepted that these models 
face problems with stability, and it has been questioned whether
such clusters are 
long-lived enough, based on evaporation and collision time-scales 
stability criteria (for a different point of view see Moffat, 1997).
Our model of Sgr A$^*$ is surprisingly simple while
 it satisfies all current observational constraints:
First, a neutrino ball is a stable object quite alike
an ordinary baryonic star, though much more massive, with
the difference that its self-gravity is compensated by the
degeneracy pressure of the neutrinos rather than thermal pressure
as in the case of a baryonic star. Second, a neutrino ball
with the abovementioned physical parameters
is compact enough as to be virtually indistinguishable
from a $2.5 \times 10^6 M_\odot$ black hole with current
observational resolution ($\approx 10^5$ Schwarzschild radii)
of the observations of proper stellar motions 
(Eckart \& Genzel, 1997; Genzel et al., 1996). Third, a neutrino
ball of this mass  can  explain its
low proper motion ($\leq$ 20 km sec$^{-1}$; Backer, 1996).
Fourth, as a bonus of our model, the neutrino ball is extended
enough to provide a much  shallower gravitational potential
than a $2.5 \times 10^6 M_\odot$ black hole for the accreting
matter, thus producing a reasonable emission flux.

\newpage
\centerline{\bf Figure captions}

Fig.1. Comparison of theoretical and observed spectra of Sgr A$^*$.
The thick solid line corresponds to the case of a neutrino ball 
of total mass $2.5 \times 10^6 M_\odot$ with
$\dot m=4 \times 10^{-3}$, while the 
thin solid line represents $\dot m= 10^{-4}$. The short-dashed line
describes the calculation with a $2.5 \times 10^6 M_\odot$  black hole,
with $\dot m= 10^{-4}$ and an accretion disk extending from 3 to
$10^5$ Schwarzschild radii. The long-dashed line corresponds to the case
when $\dot m$ is artificially reduced to $10^{-9}$.
Data points in the $<40$ GHz region are upper bounds. Note, that the
thick solid line fits the most reliable data points with the error bars.

Fig. 2. Temperature of the accretion disk as a function
of radial distance from the center. The thick line corresponds
to the case of a neutrino ball with $2.5 \times 10^6 M_\odot$
and $\dot m= 4 \times 10^{-3}$, whereas, the thin line corresponds
to a black hole with the same mass and accretion rate.
The unit of length is $r_n=2.14934 \times 10^{16}$ cm.
Note that for values larger than  3.65375, 
which corresponds to the radius of the neutrino ball, the back
hole and neutrino ball lines do overlap because the potentials are
equal.
\bsp
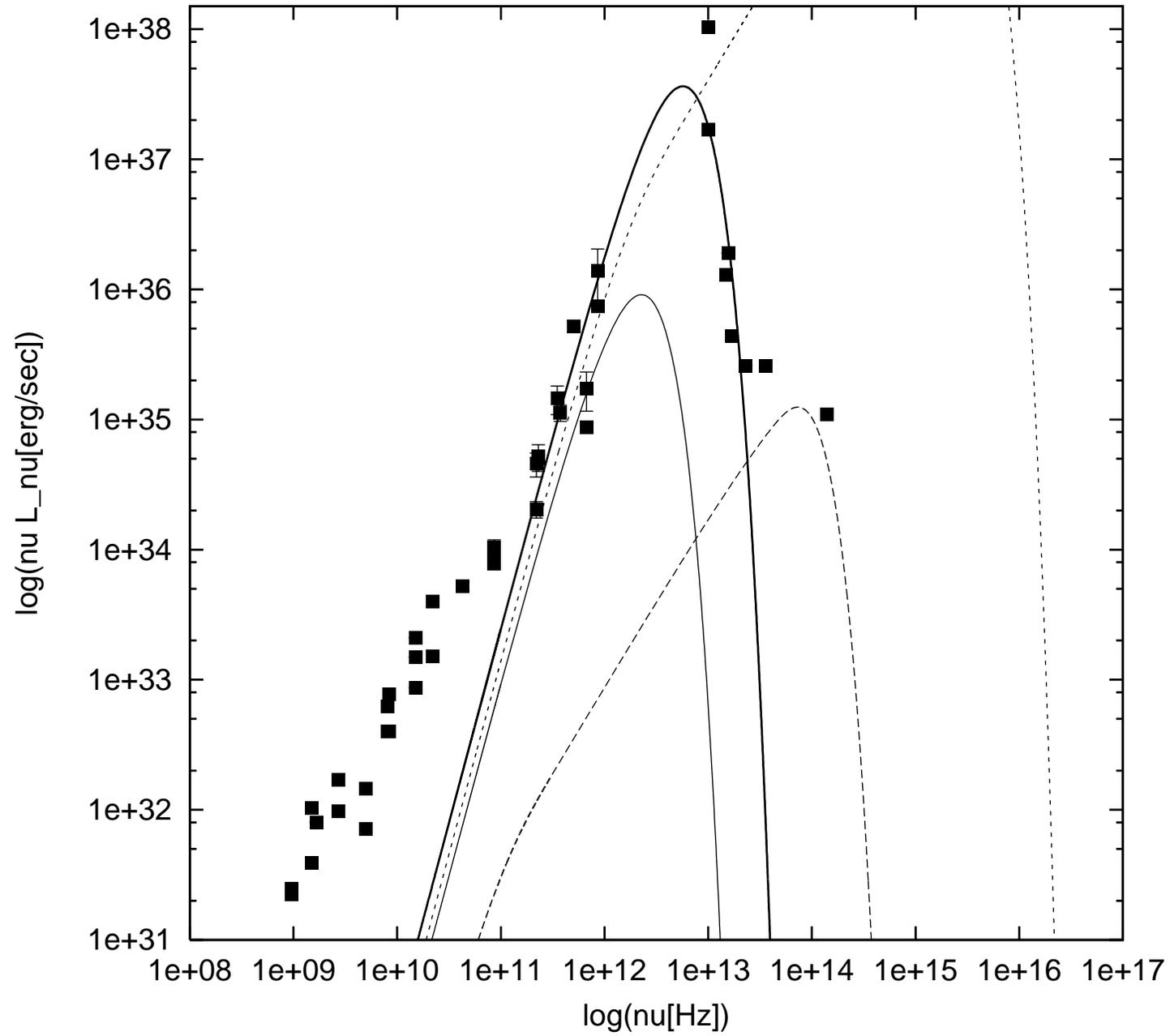
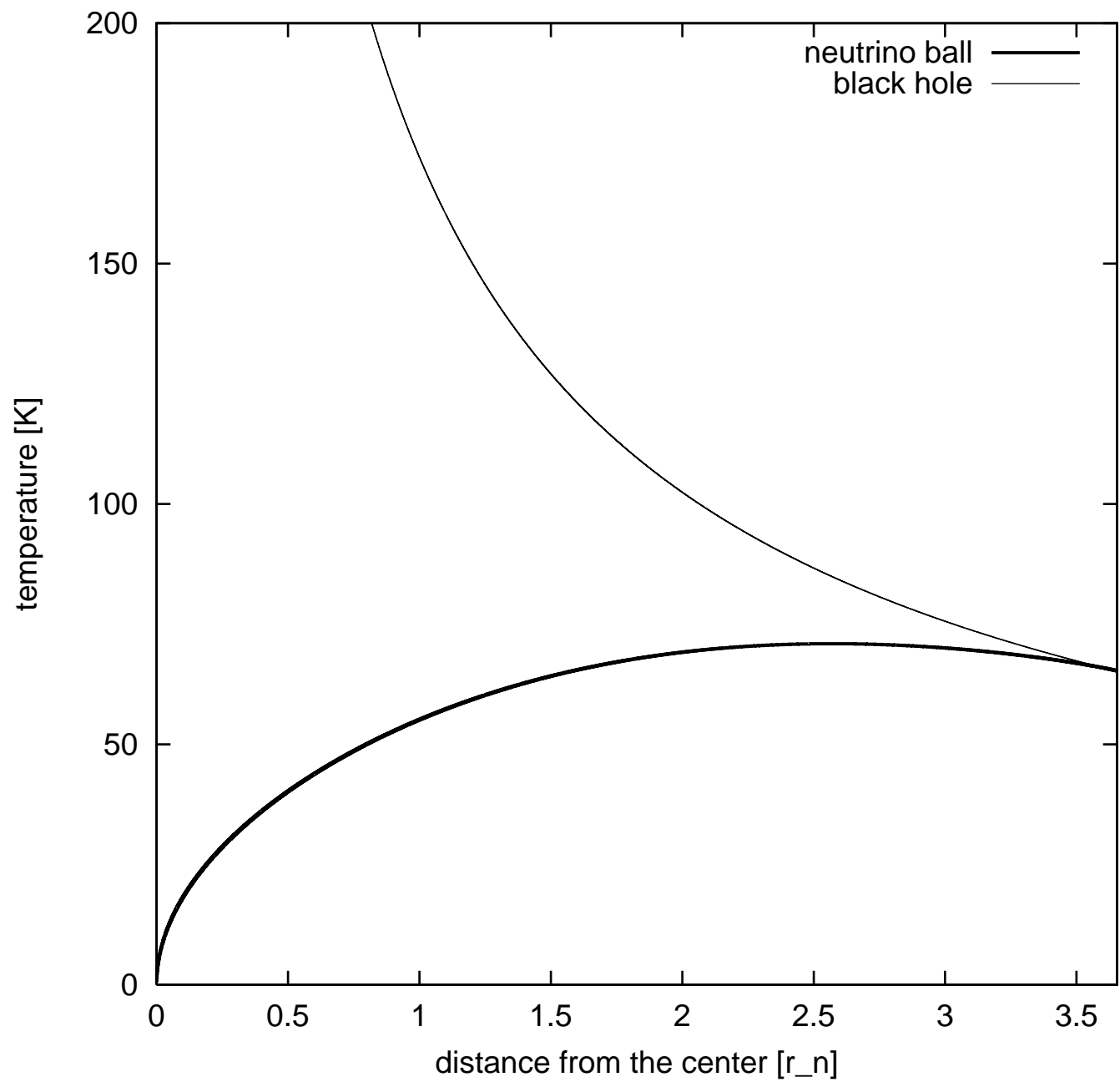
\end{document}